\documentclass[prd,aps,floats,a4paper,twocolumn,tightenlines,notitlepage]{revtex4}

\usepackage{epsfig}

\usepackage{amsmath}
\usepackage{amssymb}

\begin{document}
%
%==============================================================================

%
% Definitions for this paper.
%

\def\lap{\nabla^2}

\def\gtt{G^{t}_{~t}}
\def\grr{G^{r}_{~r}}
\def\grz{G^{r}_{~z}}
\def\gzz{G^{z}_{~z}}
\def\gaa{G^{\theta}_{~\theta}}
\def\grrzz{(G^{r}_{~r}+G^{z}_{~z})}
\def\crz{G^{r}_{~z}}
\def\crrzz{(G^{r}_{~r}-G^{z}_{~z})}
\def\IS{{\bf S}}

\def\figuremode{\small}

%
% Text of paper
%

\title{Evidence that highly non-uniform black strings have a conical waist}

\author{Barak Kol}

\email{barak_kol@phys.huji.ac.il}

\affiliation{ The Racah Institute of Physics, Hebrew University of
  Jerusalem 91904 Israel }

\author{Toby Wiseman}

\email{T.A.J.Wiseman@damtp.cam.ac.uk}

\affiliation{ DAMTP, CMS, University of Cambridge, Wilberforce Road,
  Cambridge CB3 0WA, UK }

\date{April 2003}

%==============================================================================
%
\begin{abstract}
%
%==============================================================================

  Numerical methods have allowed the construction of vacuum
  non-uniform strings. For sufficient non-uniformity, the local
  geometry about the minimal horizon sphere (the ``waist'') was
  conjectured to be a cone metric.  We are able to test this
  conjecture explicitly giving strong evidence in favour of it.
  We also show how to extend the conjecture to weakly charged strings.

  \vspace*{.3cm} {\hfill hep-th/0304070 }

%==============================================================================
%
\end{abstract}
%
%==============================================================================

\maketitle

%==============================================================================
%
\section{Introduction and Summary}
\label{sec:intro}
%
%==============================================================================

In the presence of compact dimensions massive solutions of vacuum
General Relativity may take one of several forms including the
black-hole and the black-string. For concreteness, consider the
simplest case - $d-1$ extended space-time dimensions and a compact
dimension of radius $L$, which is denoted here by the $z$ coordinate.
Any $d \ge 5$ may be considered, and here we specialise to $d=6$ for
numerical convenience.  The transition between the black hole and
string phases, which can be defined by their distinct horizon
topologies, must depend on the sole dimensionless parameter of the
problem, namely the ratio of the typical curvature radius of the
horizon and $L$. The picture we advocate here is that black objects
with curvature radius much smaller than $L$ are expected to closely
resemble a 6d Schwarzschild black hole, with a nearly round $\IS^4$
horizon, while very massive black objects will have a horizon size
much larger than $L$ and will be extended over the $z$ axis wrapping
it completely. Such objects will have $\IS^3 \times \IS^1$ horizon
topology and will be referred to as ``black strings''.  The simplest
black string is a product of the 5d Schwarzschild solution with the
spectator $z$ coordinate line and we refer to it as the ``uniform
string''. Other string solutions which are $z$ dependent we refer to
as ``non-uniform''.  Gregory and Laflamme (GL) \cite{GL1,GL2}
discovered that neutral uniform black strings develop a perturbative
$z$-dependent instability (tachyon) for masses below a critical mass
$m(L)$. Consequently the marginally tachyonic mode generates a branch
of static solutions which we call the ``GL non-uniform strings''.
Interesting links to thermodynamic stability were explored in
\cite{Gubser_Mitra1,Gubser_Mitra2,Reall,Ross}, and the instability was
further studied in \cite{Reall2,Gregory,Kang,Gibbons_Hartnoll1}.

One of the authors (BK) \cite{Kol1} used qualitative tools to
analyse the phase structure of the static solutions. The emerging
picture consists of exactly two stable phases -- the black hole
and the (stable) uniform black string connected through the
unstable GL non-uniform solutions in a first order phase
transition. These two stable solutions have not only different
horizon topologies, but the total topology of the Euclidean
solutions is different as well.  Thus a continuous topology change
(much like the conifold transition in Calabi-Yau 3-folds) occurs,
which was argued to centre around the cone over $\IS^2 \times
\IS^{d-3}$.

This conjecture addresses two questions. Firstly it conjectures that
there is only one solution for large mass objects. This question is
clearly essential for understanding how astrophysical black holes are
manifested, and whether there are potentially observable consequences.
Secondly, based on Gubser's higher order perturbative construction of
the GL non-uniform strings \cite{Gubser}, it predicts all strings in
this branch have higher mass than the unstable uniform strings, and so
\emph{these} solutions are unlikely to be the end state for GL decay
as suggested by Horowitz and Maeda \cite{HM1}\footnote{It is important
  to understand this decay end state as a stable string will always
  lose mass by Hawking radiation until it becomes unstable.}.  For
another approach to the phase structure see \cite{Harmark_Obers}.  For
implications to black hole uniqueness and for a description of the
explosive transitions with Planck-scale power see \cite{Kol2,Kol3}.

Let us recall the reason for the appearance of the cone and some
more details. One considers a highly non-uniform string as well as
a large black hole (a black hole which hardly fits in the compact
dimension) and focuses on the vicinity of the ``waist'' -- the
minimal sphere in the black string, and on the ``gap'' between
``north'' and ``south'' poles in the black hole. Then one finds
that the topology at some distance away from the waist is $\IS^2
\times \IS^{d-3}$ where the second sphere comes simply from the
spherical symmetry and the first includes the Euclidean time.
Moreover the spatial $\IS^{d-3}$ is topologically trivial in the
black hole (it can shrink) and correspondingly the $\IS^2$ for the
black string. Hence a topology change analogous to the ``pyramid
picture'' of the conifold suggests itself where the cone here is
over the metrically round $\IS^2 \times \IS^{d-3}$.  The isometry
of the cone is enhanced (relative to the rest of the solution) and
in particular the geometry near the waist depends on only one
variable -- the distance $\rho$ from the waist. Not only does this
enhanced symmetry realize the topology change in a simple setting,
but the equations of motion actually forbid some more general
ansatze which were tried.

Analytic solutions are not available, despite some attempts
\cite{deSmet,Harmark_Obers} and some relevant constructions in 4-d
\cite{Myers,Korotkin_Nicolai,Frolov} (where the proper size of the
compact dimension does not asymptote to a constant, and there are
no black string solutions), which do not generalise to $d>4$
\cite{Emparan_Reall1}. One of the authors (TW)
\cite{Wiseman1,Wiseman2} showed that in general the numerical
method of relaxation could be applied to gravitostatics by
overcoming the conceptual issues (largely related to imposing the
constraint equations and boundary conditions) and in particular
was able to trace numerically the GL branch of non-uniform strings
in the case $d = 6$.  Large amounts of non-uniformity were reached
($R_{\mbox{max}}/ R_{\mbox{min}} \sim 9$) exposing the asymptotic
behaviour for the GL non-uniform branch including limiting mass
and temperature. Moreover, it was found that the whole branch has
a mass higher than the critical string and therefore cannot serve
as an endpoint for decay of unstable uniform strings. The
agreement with the theoretical predictions of \cite{Kol1} were
further strengthened in \cite{Wiseman3} where global properties of
the solutions were studied and shown to behave in accord with the
conjecture.

In this paper we perform a much stronger test of the conjecture in
\cite{Kol1} by comparing the numerical geometries with the predicted
cone. A summary of the method and results follows.

First one is required to identify the transformation from cone
coordinates to the numerical metric, and then to compare the two
metrics. The cleanest coordinate to extract is $\rho$, and the
``smoking gun'' test involves comparing the numerical Kretschmann
curvature invariant $K_{num}$ with the cone prediction, $K_{cone}$,
which due to the symmetry properties of the cone, only depends on
$\rho$. A quantitative test shows that over a region where $K$ changes
by orders of magnitude agreement is found and $K_{num}/K_{cone}$
changes only by a factor of about two, which we consider to be a
sub-leading effect \footnote{We attribute the mismatch to the local
  nature of the cone approximation and the additional issues mentioned
  at the end of the paragraph.}.  Next we infer the azimuthal angle
$\chi$ from the numerical solution.  Actually, comparison of the
metric gives $\sin(\chi)$ and so we must test that the maximum of the
expression for $\sin(\chi)$ is consistent with 1, and we see this is
the case.  Three metric functions remain to be compared, and indeed
show good agreement. This agreement is particularly satisfying given
the low resolution of the grid over the relevant region, and the
finite value of non-uniformity which can be achieved.

In summary, we conclude that the numerical 6-d \footnote{It would be
  interesting to check numerically the prediction of a critical
  dimension $d_c=10$ \cite{Kol1}.  For $d<10$ the cone is unstable
  while it is stable for $d>10$ (see also
  \cite{Bohm,Gibbons_Hartnoll1,Herzog_Klebanov}). Thus there is the
  possibility that for $d>10$ a large black hole will be stable up to
  the point where its horizon self-intersects, whereas for lower
  dimensions a tachyonic instability must kick in earlier. }
geometry at the vicinity of the waist reveals what is essentially
a cone structure, supporting the analysis of \cite{Kol1}.
Sub-leading deviations from the cone remain and may hold further
clues and refinements. We conclude with an argument that the waist
cone geometry remains a good approximation if the string carries a
small electric charge, and construct the behaviour of the electric
potential near the cone tip.

%==============================================================================
%
\section{Numerical string solutions}
\label{sec:numerics}
%
%==============================================================================

Following Gubser, we define,
\begin{equation}
\lambda = \frac{1}{2} \left( \frac{R_{max}}{R_{min}} - 1 \right)
\end{equation}
where $R_{max, min}$ are the maximum and minimum sphere radii in the
horizon. Non-uniform strings were constructed non-linearly in
$\lambda$ in \cite{Wiseman2} using the methods developed in
\cite{Wiseman1}. For technical reasons, the computation was performed
in 6 dimensions, although it was shown that 6-dimensional strings have
the same qualitative thermodynamic behaviour as 5 dimensional strings
to leading order in $\lambda$.  The numerical method uses the metric,
\begin{equation}
ds^2 = \frac{r^2}{m + r^2} e^{2 A} d\tau^2 + e^{2 B} ( dr^2 + dz^2 ) + e^{2 C}
( m + r^2 ) d\Omega^2_{3}
\label{eq:num_metric}
\end{equation}
where $A, B, C$ are functions of $r, z$ and $z = [0,L]$ is the
periodic coordinate of the $S^1$ on which the string wraps. Due to
reflection symmetry we actually consider the half period $z = [0,
L_n]$ with $L_n = L/2$. We have Euclideanised time and now the
usual Euclidean constraint makes $\tau$ periodic, $\tau = [0,
\tau_{max}]$ with,
\begin{equation}
\tau_{max} = 2 \pi \sqrt{m} e^{(B-A)} \mid_{r=0} = \frac{1}{T}
\end{equation}
with $T$ the horizon temperature of the string. Furthermore $A, B, C
\rightarrow 0$ as $r \rightarrow \infty$ is chosen.  Choosing units so
that $m = 1$, we take $L = L_{crit}$ as a convenient initial condition
for the relaxation, so $L_n = 2.476$ and $A, B, C$ decay exponentially
for small $\lambda$. The metric components are then solved by
relaxation, and $\lambda \simeq 4$ is attainable. The method
unfortunately is limited by resolution, and to proceed to higher
$\lambda$ would require much computation time, and we suspect that it
is more fruitful to attempt to improve the method itself before more
detailed calculations are performed.

For large $\lambda$ the waist shrinks, compatible with vanishing as
$\lambda \rightarrow \infty$, whilst the maximal sphere radius
$R_{max}$ appears to asymptote to a constant radius \cite{Wiseman3}.
Other global properties of the solution were also measured to check
compatibility with the black string/hole transition conjecture, namely
that the horizon temperature, mass, and the proper distance along the
horizon appear to asymptote to constants for large $\lambda$.
Assuming the conjecture to be true, this is intuitively explained by
the fact that at large $\lambda$ the geometry hardly changes with
varying $\lambda$, except near the very cone tip. As this region is
small, it is only the very massive Kaluza-Klein states that
participate, and these are strongly exponentially damped away from the
waist, and therefore do not contribute much to variation of global
properties.

As in any numerical GR context, it is difficult to assess the physical
significance of constraint violations. For small $\lambda$ we may
compare non-linear solutions with Gubser's perturbation theory method.
However, for large $\lambda$ where we become resolution limited, it is
harder to assess error. In \cite{Wiseman2} this was done by direct
observation of the constraints, and furthermore using the first law to
compare mass asymptotically measured with that integrated from the
horizon geometry. Good agreement was found, but any additional tests
of numerical accuracy are obviously important.  The cone conjecture is
a perfect example of such a test; at large $\lambda$, if the geometry
is indeed a cone at the waist, we may explicitly check that the
numerical metric does indeed reproduce this. That is exactly what we
do here. Furthermore, if the conjecture is correct it may allow
numerical methods to start relaxation about the cone geometry, for
either the black string or black hole branch, and therefore improve
numerical stability.

%==============================================================================
%
\section{Testing the cone}
\label{sec:cone}
%
%==============================================================================

In 6-dimensions we may now write the Euclidean cone metric over the
base $S^2 \times S^3$,
\begin{equation}
ds^2 =  d\rho^2 + \rho^2 \left( \frac{1}{4} \left( d\chi^2 + \sin^2{\chi} N^2 d\tau^2 \right) + \frac{1}{2} d\Omega^2_{3} \right)
\label{eq:cone_metric}
\end{equation}
where $\tau, \Omega_{3}$ are the same coordinates as for the Euclidean
string \eqref{eq:num_metric}. For the metric to be smooth away from
the apex, $\chi$ is an angle coordinate with range $\chi = (0, \pi)$.
Furthermore, $N$ must be chosen as $2 \pi / \tau_{max}$ to ensure $N
d\tau$ behaves correctly as an angle in the $S^2$.  We note, as
emphasised in \cite{Kol1}, that there are no physical free parameters
once one has postulated this cone base.

We may now directly compare the numerical metric \eqref{eq:num_metric}
and the cone \eqref{eq:cone_metric}. Comparison of $dt^2,
d\Omega^2_{3}$ give,
\begin{eqnarray}
\rho(r,z) & = & e^{C(r,z)} \sqrt{2 (1 + r^2)}
\nonumber \\
\sin{\chi(r,z)} & = & \frac{r}{N \sqrt{2} (1 + r^2)} e^{A(r,z)-C(r,z)}
\label{eq:rho_chi}
\end{eqnarray}
Thus $\rho$ and $\chi$ are completely determined and there is no
residual coordinate freedom to complicate the comparison. Knowing
$\rho$ and $\chi$, we may simply compute the remaining components of
the metric \eqref{eq:cone_metric} and explicitly compare to
\eqref{eq:num_metric}. Then,
\begin{eqnarray}
e^{2 B(r,z)} ( dr^2 + dz^2 ) & = & d\rho(r,z)^2 + \frac{1}{4} \rho(r,z)^2 d\chi(r,z)^2
\nonumber \\
& = & c1\, dr^2 + c2\, dz^2 + 2\, c3\, dr dz
\label{eq:rzmetric}
\end{eqnarray}
where
\begin{eqnarray}
c1 & = & (\partial_r \rho)^2 + \frac{1}{4} \rho^2 (\partial_r \chi)^2
\nonumber \\
c2 & = & (\partial_z \rho)^2 + \frac{1}{4} \rho^2 (\partial_z \chi)^2
\nonumber \\
c3 & = & (\partial_r \rho) (\partial_z \rho) + \frac{1}{4} \rho^2 (\partial_r \chi) (\partial_z \chi))
\end{eqnarray}

There are two complicating factors in executing this comparison.
Firstly we cannot reach $\lambda \rightarrow \infty$ but only $\lambda
\simeq 4$ and thus we do not expect perfect agreement with the cone,
as for finite $\lambda$ the tip will be missing.  Secondly, numerical
resolution is limited, and we only expect good agreement in the region
near the waist. For the maximum $\lambda$ computed the $r, z$ lattice
resolution is $dr = 0.025$ and $dz = 0.026$. We will find good
agreement with the cone is a region of coordinate width $\Delta r
\simeq \Delta z \sim 0.5$, restricting us to a 20*20 sub-grid of the
whole lattice. Considering large gradients are present, it is unclear
how satisfactory this resolution is \footnote{Bear in mind that we
  cannot compare with a lower resolution as this does not allow
  $\lambda$ as large as this to be computed.  Furthermore we are
  unable to compute higher resolutions in a reasonable time.}.  Both
of these factors add considerable uncertainty in assessing agreement.
However, we will see that the agreement is sufficient that the cone is
clearly a good approximation, even though it is difficult to make the
statement more precise at this time.

\subsection{Comparisons of curvature invariant}

\begin{figure}[t]
\psfig{file=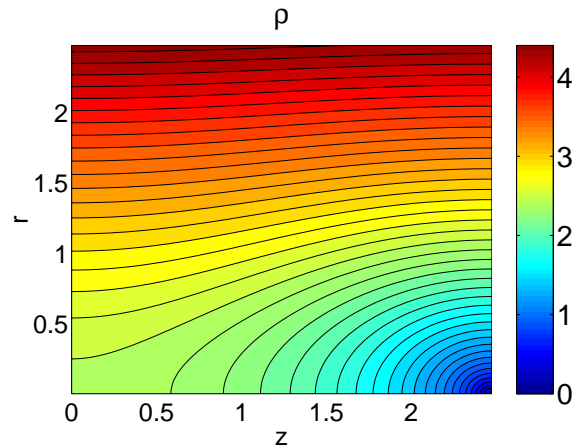,width=3in}
\caption{ Plot of $\rho$ measured from the $\lambda = 3.9$ string solution.
\label{fig:rho}
}
\end{figure}

\begin{figure}[t]
  \psfig{file=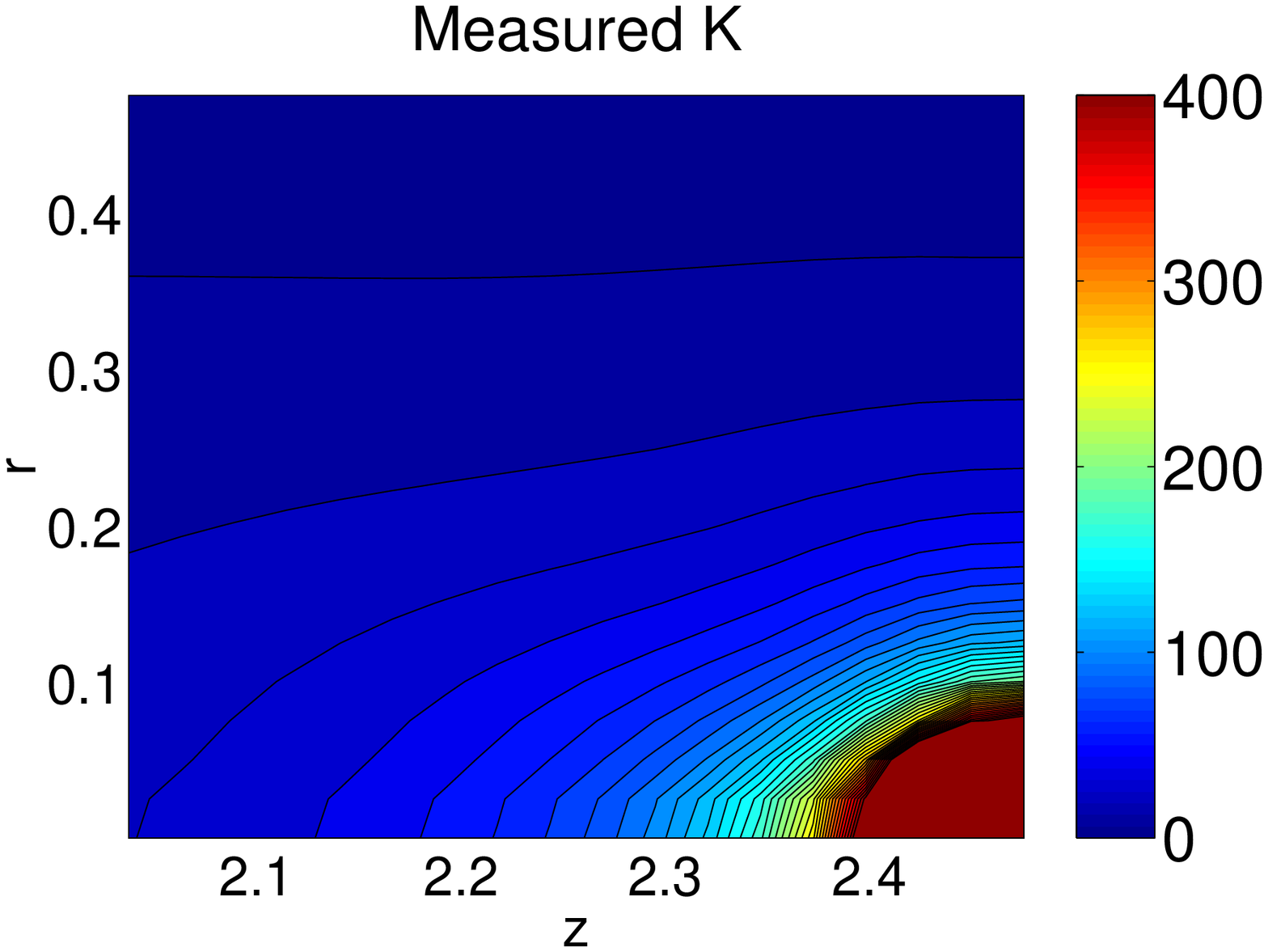,width=3in}
  \psfig{file=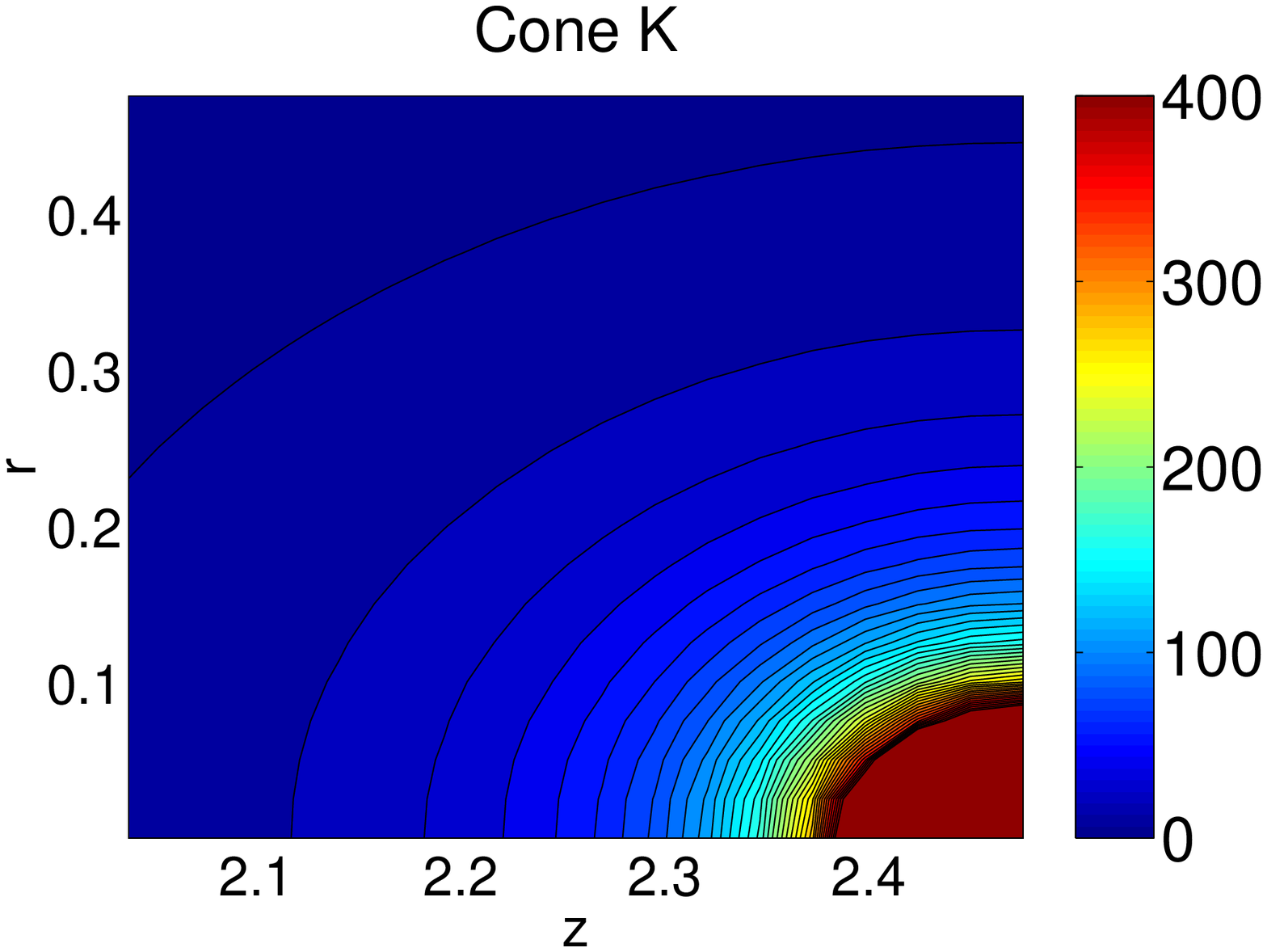,width=3in}
  \psfig{file=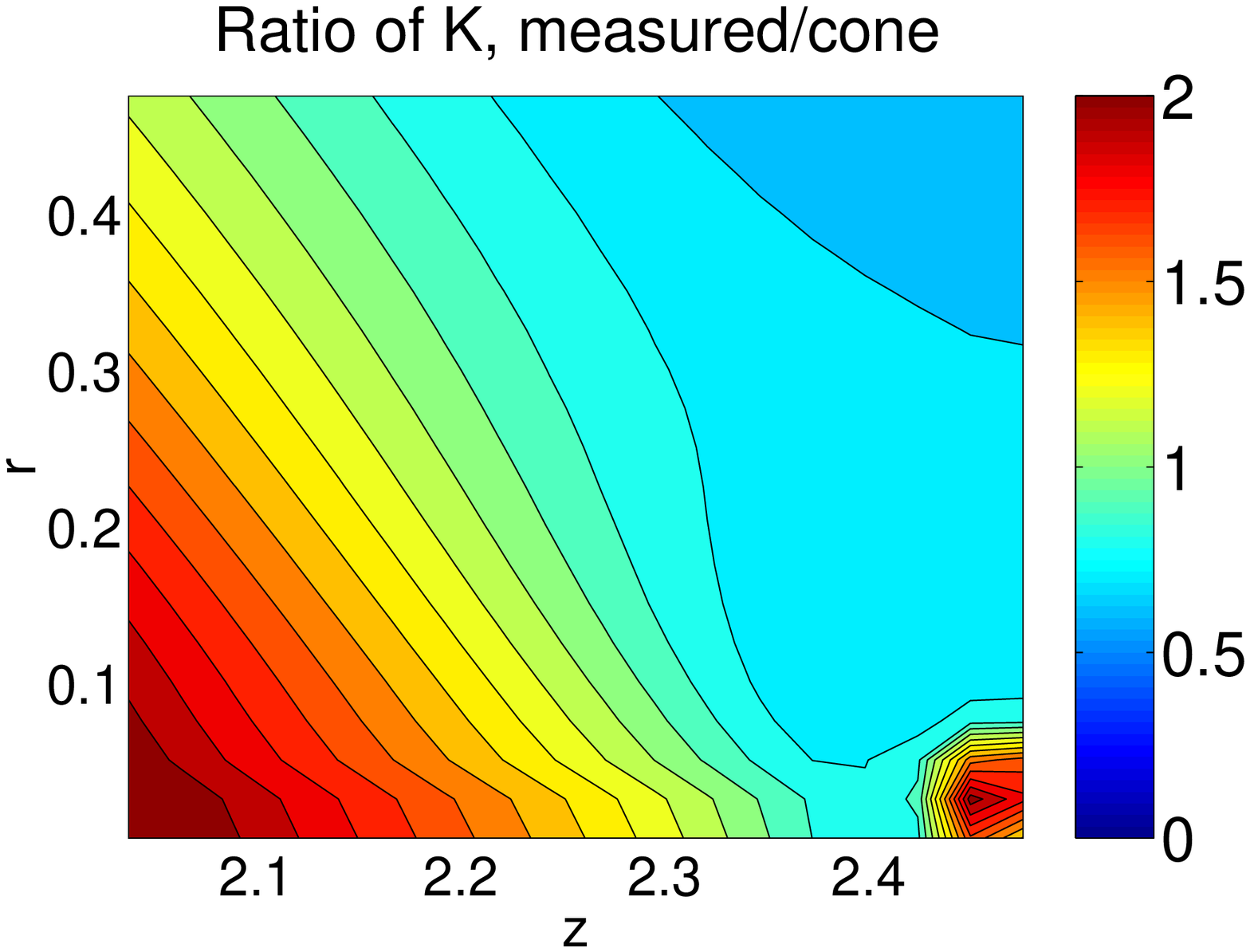,width=3in}
\caption{ Plot showing $K$, the scalar curvature invariant, measured from the string ($\lambda = 3.9$) compared to the cone prediction. The ratio is also shown, and we see good agreement slightly away from the apex, where finite $\lambda$ `resolves' the cone. Moving far from the cone ie. $z<2.1$ or $r>0.4$ the cone is no longer a good approximation, the actual curvature becoming more homogeneous in $z$.
\label{fig:K}
}
\end{figure}

In figure \ref{fig:rho} we plot the function $\rho$ against $r,z$ for
the most non-uniform solution found numerically, with $\lambda = 3.9$.
We see very clearly the correct qualitative behaviour of $\rho$ for a
cone near to the minimal sphere at $r = 0, z = L_n$, where $\rho$ has
its minimum. The degree to which the cone is `resolved' can be
estimated from this minimum value. For the value of $\lambda$ plotted
in the figure, $\rho_{min} = 0.27$. This indicates that the cone will
be a good approximation for $\rho > \rho_{min}$, but $\rho << L_n$, the
characteristic scale of the compactification.

Quantities involving only $\rho$ provide the easiest numerical
comparisons. In the next section we see that computing $\chi$ from
$A,B,C$ involves further processing, and thus tests only involving
$\rho$ should be more robust. A non-trivial test of the cone
conjecture, simply involving $\rho$ is a comparison of the Kretschmann
curvature scalar with the cone prediction.  For the cone metric
\eqref{eq:cone_metric} one finds,
\begin{equation}
K = R^{\mu\nu\alpha\beta} R_{\mu\nu\alpha\beta} = \frac{48}{\rho^4}
\end{equation}
One can compute the same quantity for the numerical solutions
directly from $A, B, C$ and the two are plotted in figure
\ref{fig:K} for the region near the waist.  We also plot the ratio
of these quantities in the same figure. We see that at the waist
agreement is lost, as we expect because we are at finite
$\lambda$. However, moving away from the apex, agreement seems to
be good, considering the relatively low numerical resolution
available over such a small subregion of the string solution. The
ratio is close to one, and we also see a possible sub-leading
behaviour in $r$, and less strongly in $z$ as we move away from
the `apex'. Moving further from the waist we see the curvature
contribution from the cone decays sufficiently that a more
homogeneous curvature emerges, as expected for a string.

\subsection{Full metric comparison}

\begin{figure}[t]
\psfig{file=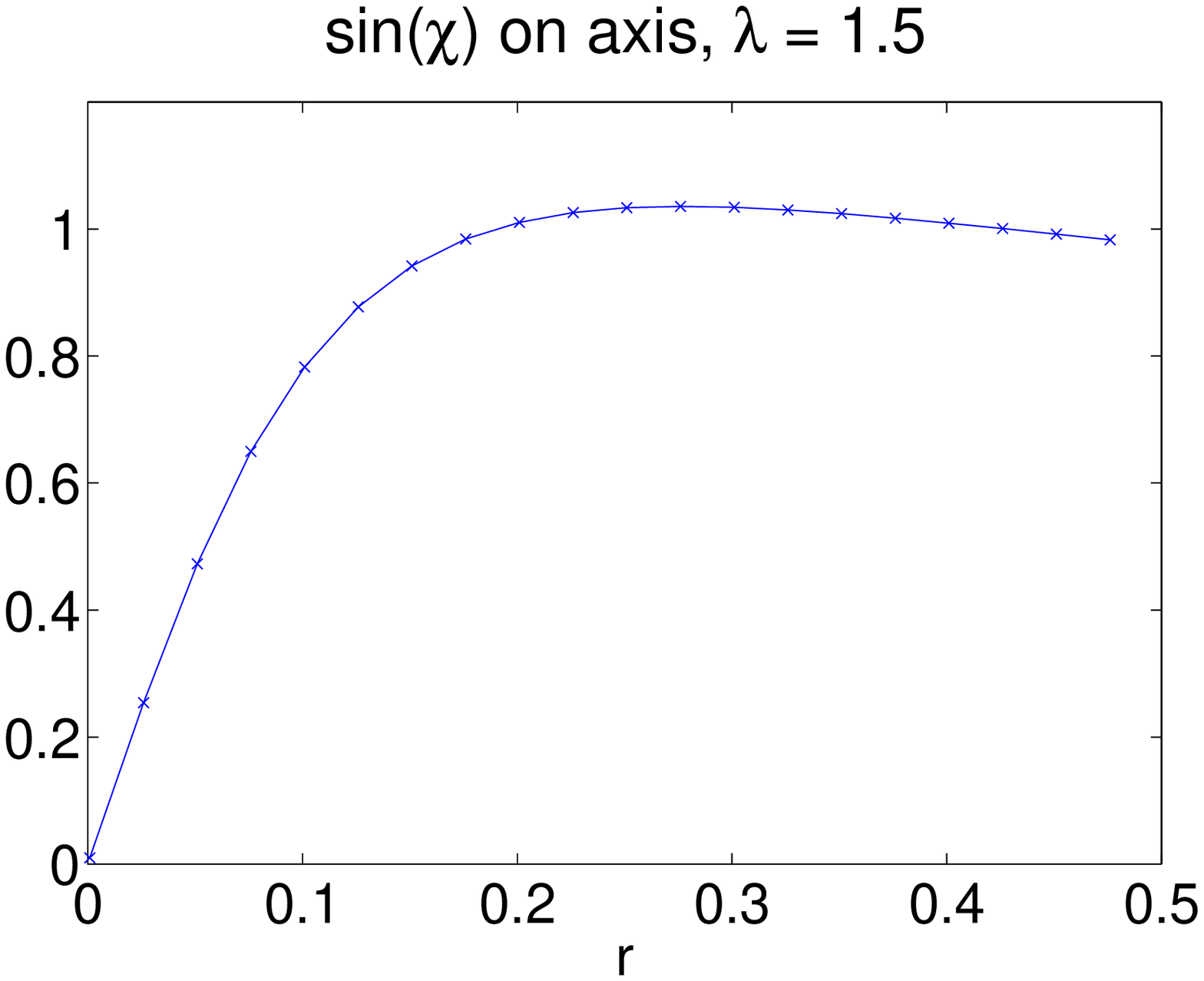,width=2.5in}
\psfig{file=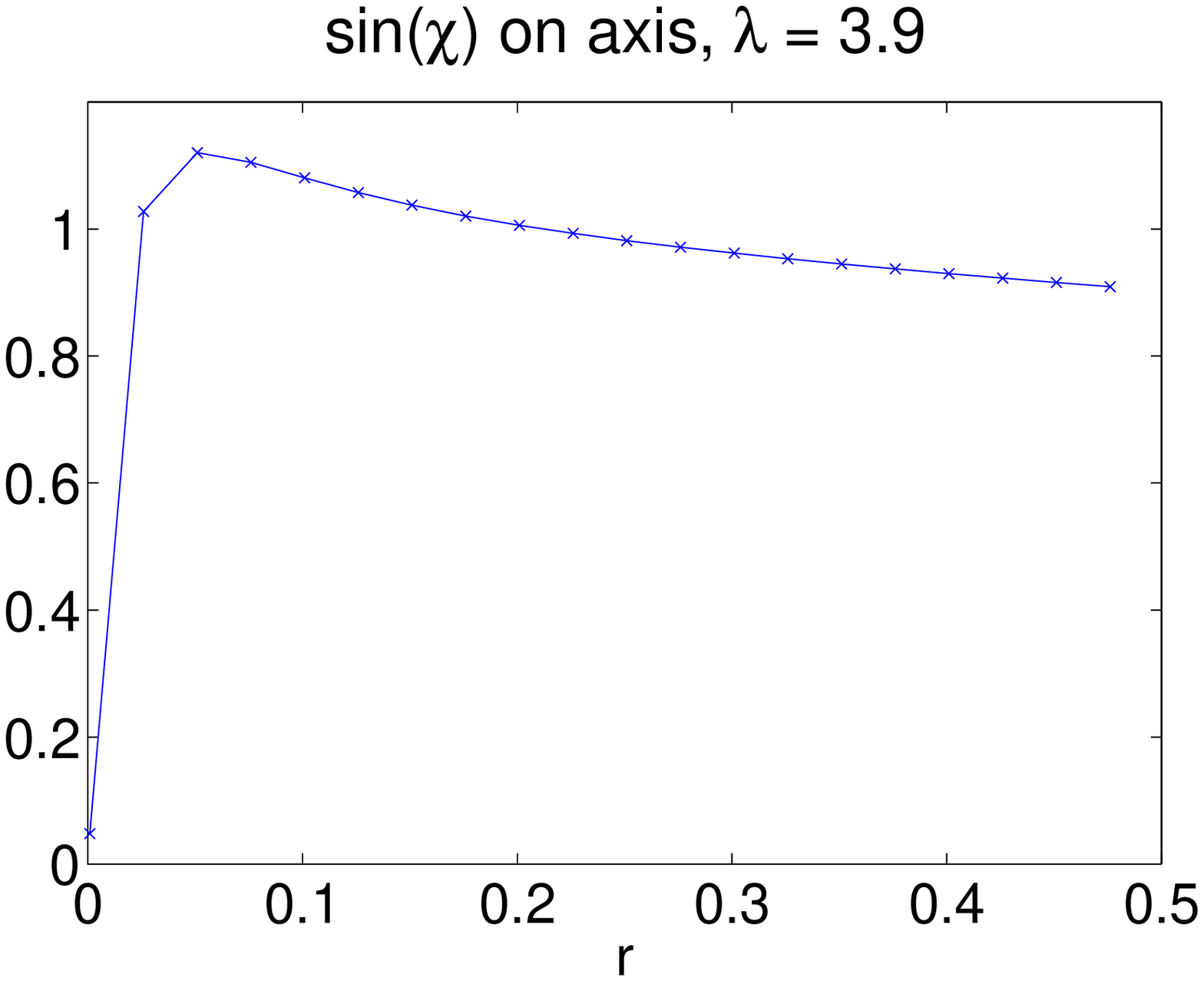,width=2.5in}
\caption{ Plots of $\sin{\chi}$ on the $z = L_n$ axis measured from string solutions with $\lambda = 1.5, 3.9$. For the larger $\lambda$ we see the curve become more `constant', and furthermore, the value is remarkably consistent with one, the cone prediction.
\label{fig:chimax}
}
\end{figure}

\begin{figure}[t]
\psfig{file=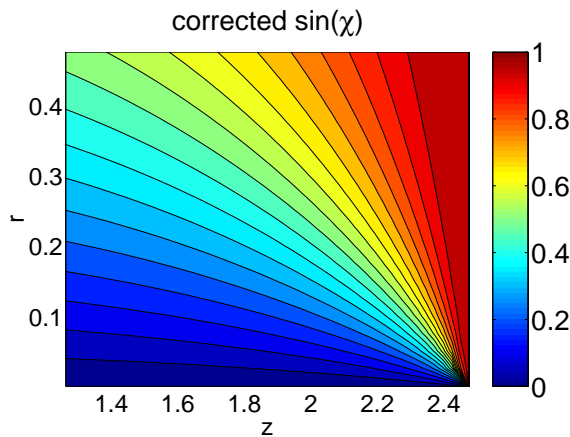,width=3in}
\caption{ Plot of $\sin{\chi}$ computed from $\lambda = 3.9$ string, including normalisation correction to ensure $\sin{\chi} = 1$ at $z = L_n$, so that $\chi$ can be meaningfully extracted.
\label{fig:sinchi}
}
\end{figure}

\begin{figure}[t]
  \psfig{file=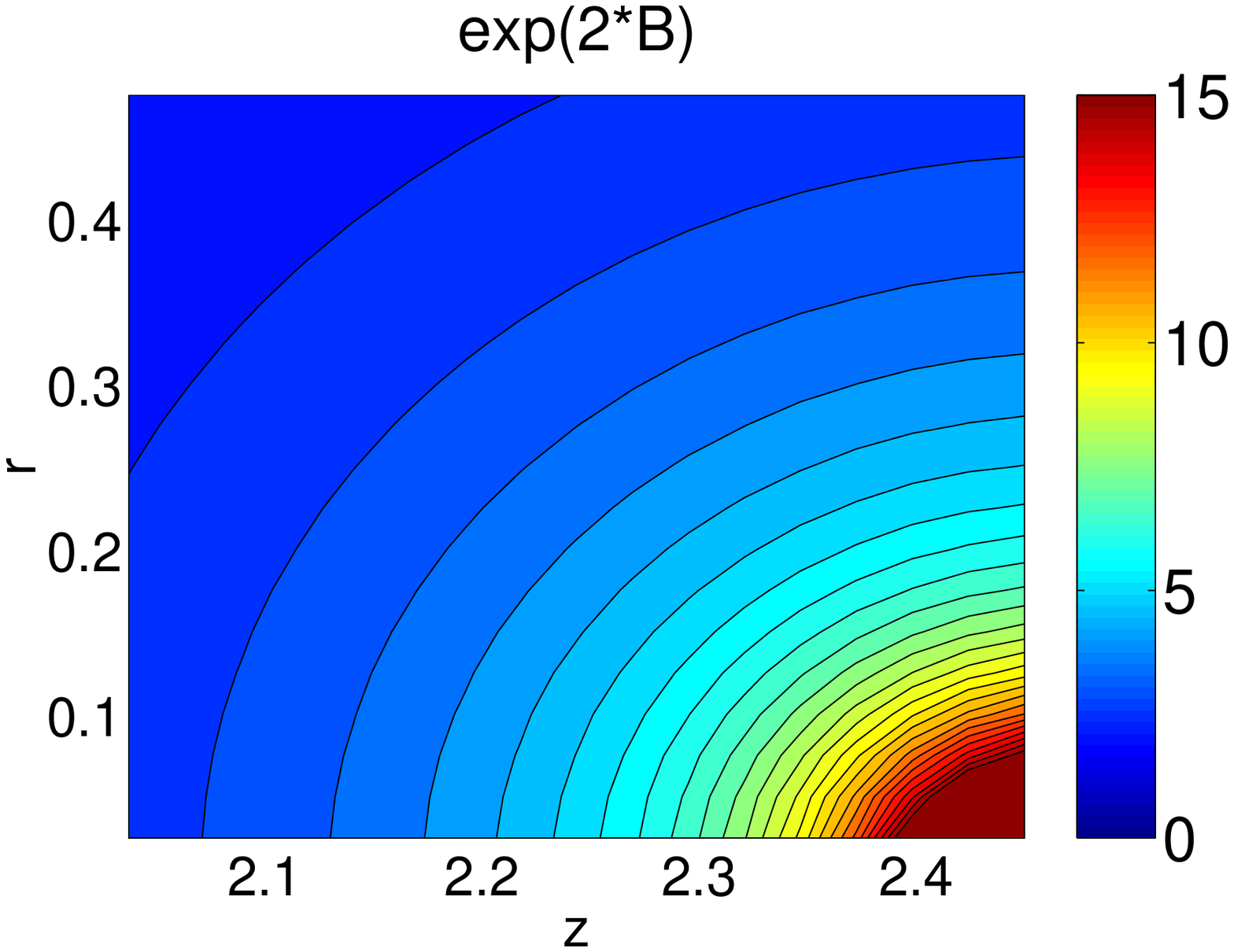,width=2.8in}
  \psfig{file=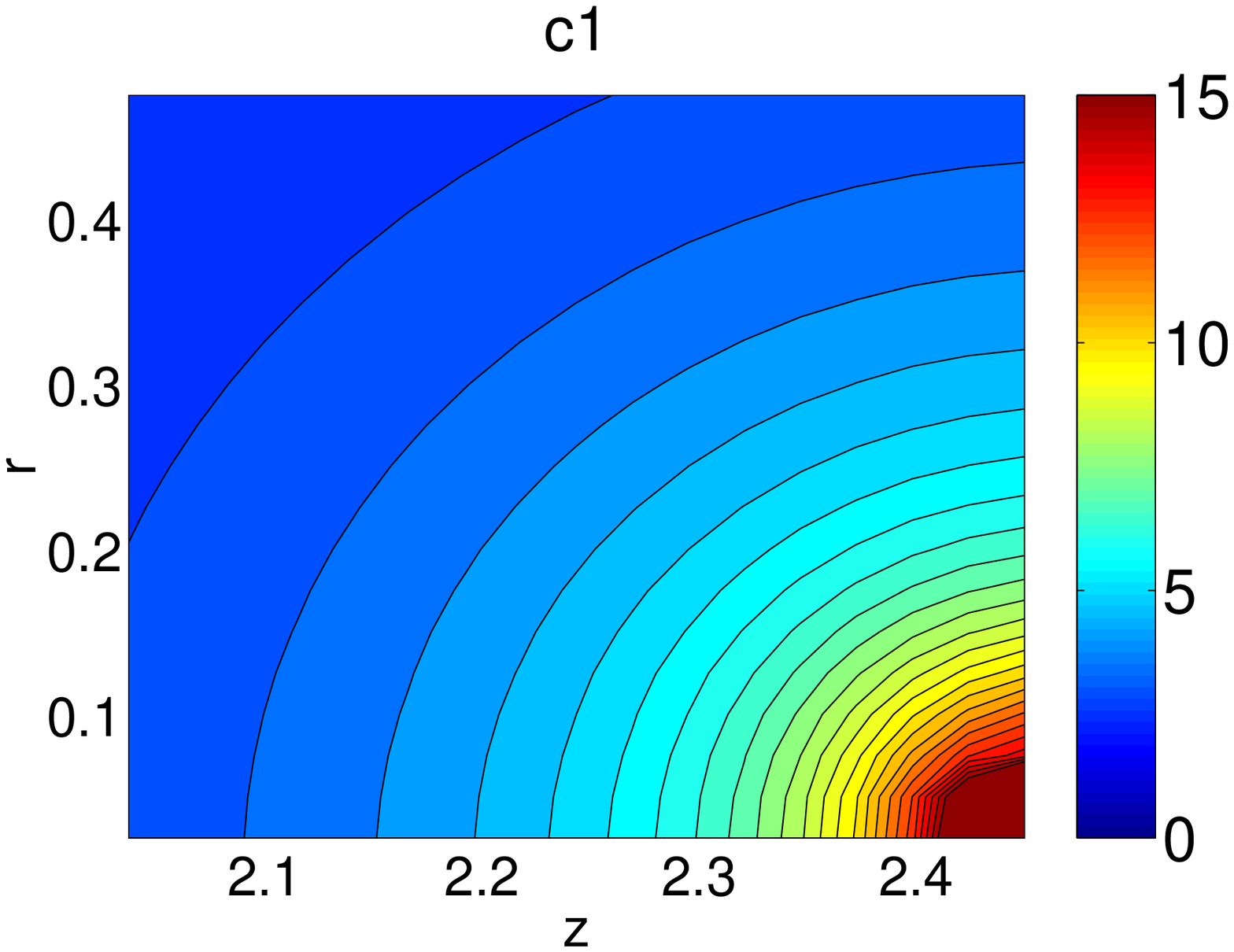,width=2.8in}
  \psfig{file=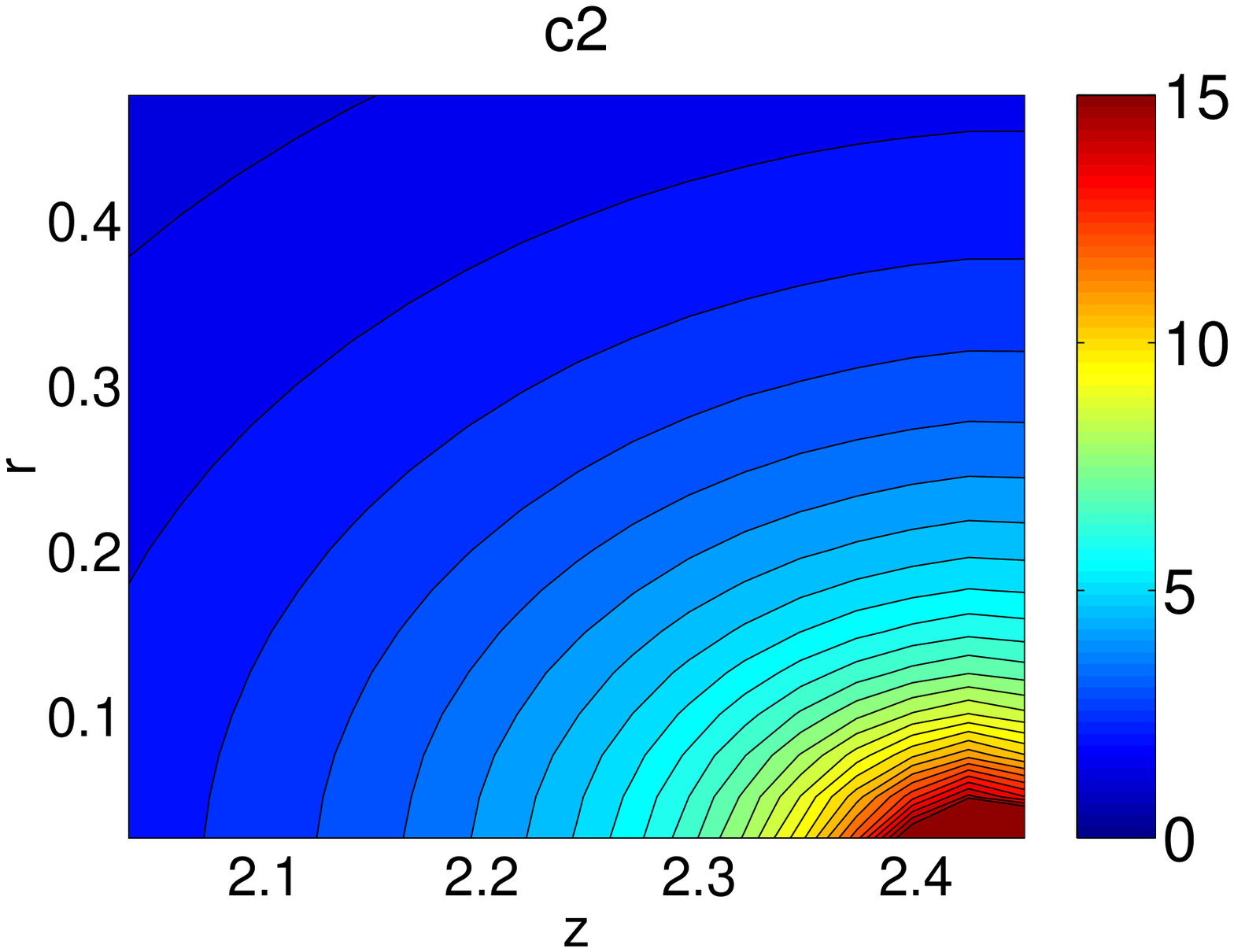,width=2.8in}
\caption{ Plots of remaining metric components $c1, c2$ computed from the numerical solution with $\lambda = 3.9$ using the `corrected' $\sin{\chi}$. The cone predicts that $c1 = c2 = e^{2 B}$ which is also plotted for comparison. Very good agreement is seen, given that the underlying data is on a lattice of only approximately $20*20$ and resolution effects may give significant numerical errors.
\label{fig:c1c2}
}
\end{figure}

\begin{figure}[t]
\psfig{file=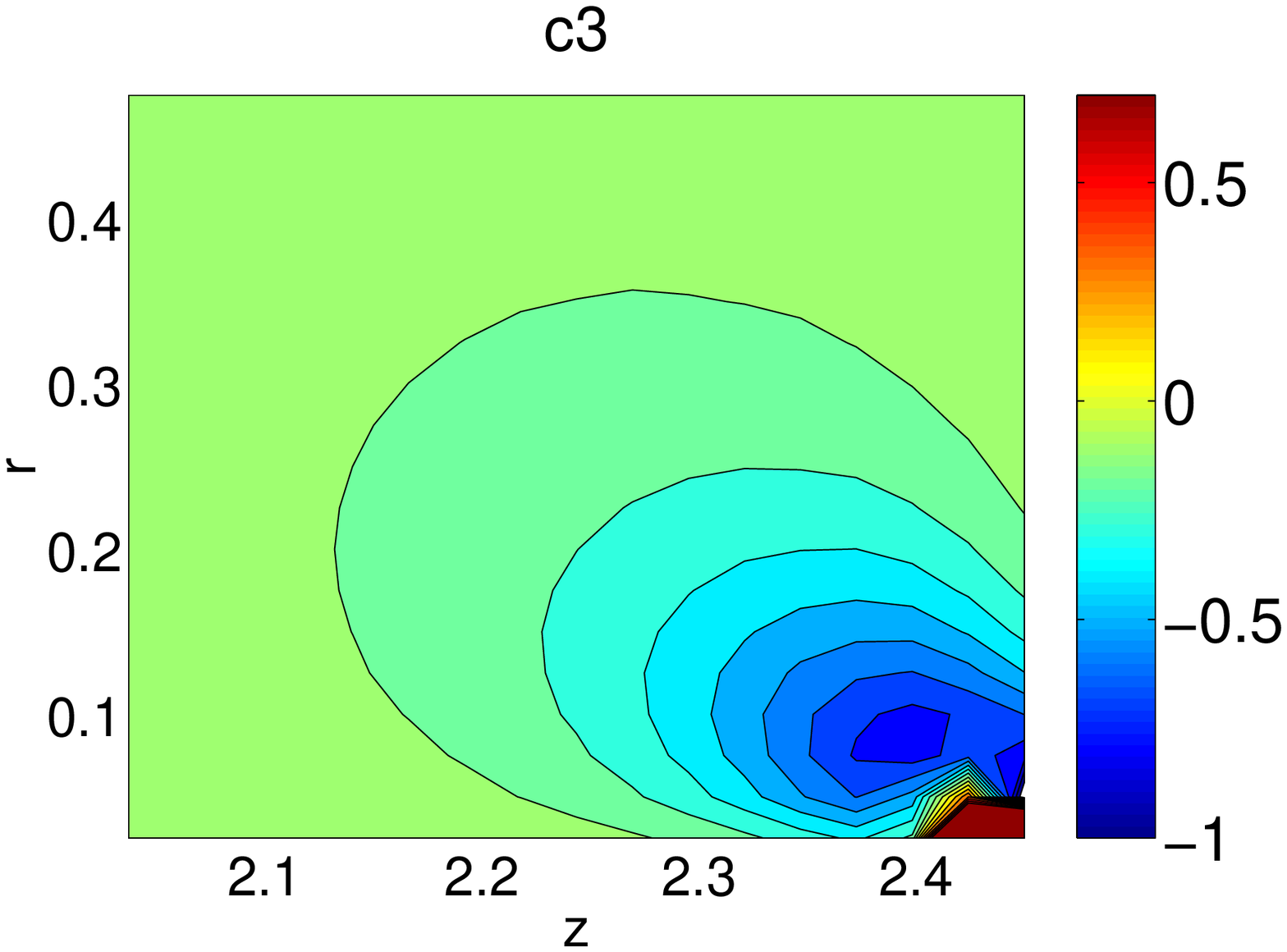,width=3in}
\caption{
  The off-diagonal metric component $c3$ which the cone predicts to be
  zero. Away from the resolved apex, we see it is indeed much smaller
  than the values of $c1, c2$ in the previous plot, again being
  consistent with the cone ansatz.
\label{fig:c3}
}
\end{figure}

This curvature invariant indicates the size of the region where we
expect good agreement. Now we perform a full metric comparison using
$\chi$ as well as $\rho$. However, use of $\chi$ is subtle. For the
cone obviously $\sin{\chi} = 1$ for $\chi = \pi/2$. However, we are
measuring $\sin{\chi}$ directly, not $\chi$, and due to finite
$\lambda$ resolving of the cone, and limited numerical resolution, we
no longer can expect that $\sin{\chi} = 1$ exactly at $z = L_n$. In
figure \ref{fig:chimax} we plot $\sin{\chi}$ for two values of
$\lambda = 1.5, 3.9$ at $z = L_n$, over the range of $r$ indicated in
the previous section to yield good agreement. Not only do we clearly
see that for the larger $\lambda$ value $\sin{\chi}$ on the axis $z =
L_n$ appears to be more constant, but moreover this constant is indeed
consistent with one.  Bearing in mind the low resolution we regard
this as remarkably good agreement. For the lower value of $\lambda$, a
constant behaviour is still seen at large $r$, again consistent with
one, but there is a larger region near the apex where $\sin{\chi}$ is
clearly not a constant, being approximately linear in $r$.

In order to compute \eqref{eq:rzmetric} we must compute $\chi$ from
the $\sin{\chi}$ and obviously this may lead to pathologies, as we see
above that the $\sin{\chi}$ measured at $z = L_n$ is larger than one
at its peak. Therefore, we normalise the $\sin{\chi(r,z)}$ measured by
its value on the $z = L_n$ axis (which `should' be one, and we have
seen does well approximate one) as,
\begin{eqnarray}
\sin{\chi(r,z)}_{corrected} & = &\sin{\chi(r,z)} / \sin{\chi(r_0,L_n)}
\nonumber \\
\rho(r_0,L_n) & = & \rho(r,z)
\end{eqnarray}
so we use contours of $\rho$ to find the value on the axis $z = L_n$ to
normalise by. This tidies up $\chi$ at $z = L_n$, and it is difficult to
see how to extract $\chi$ without dealing with this issue.

Now we plot $\sin{\chi}_{corrected}$ in figure \ref{fig:sinchi}. We
see exactly the type of behaviour we would expect, with contours of
$\chi$ parameterising an angle near the apex.  Then in figure
\ref{fig:c1c2} we plot the $dr^2$ and $dz^2$ components of the metric
($c1, c2$) \eqref{eq:rzmetric} calculated using the corrected $\chi$.
According to the cone conjecture, these components should be equal to
each other, and to $e^{2 B(r,z)}$, which we also plot in this figure.
We see rather remarkable consistency with the cone prediction. The
remaining test is that the measured $dr dz$ component ($c3$) should be
zero, or numerically speaking, much smaller than the $dr^2, dz^2$
components.  This is finally plotted in figure \ref{fig:c3}, and again
we see that the prediction is indeed true.

%==============================================================================
%
\section{Charged strings}
\label{sec:charge}
%
%==============================================================================

Uniform strings electrically charged under a Maxwell field can be
constructed from black hole solutions of Einstein-Maxwell-Dilaton
theory \cite{GHT,Gibbons_Maeda,Garfinkle}.  For
small charges non-uniform deformations will exist since the GL
instability persists.  As the string horizon is an equipotential
surface which is ``dented'' we expect the field to be reduced in the
waist region -- this is the complementary effect to the appearance of
high fields near a ``pointed protrusion''.  Here we consider whether
we can quantify this further, using our concrete example $d = 6$.

Unfortunately at this time we know of no generalisation of the cone to
include an electric flux. Therefore we simply consider putting a small
charge on the cone and neglecting its back-reaction. Making the
electric ansatz that the vector potential $A_t = \phi(\rho, \chi)$
with other components being zero, one finds,
\begin{equation}
\partial_{\rho}^2 \phi + \frac{3}{\rho} \partial_\rho \phi + \frac{4}
 {\rho^2} \left( \partial_\chi^2 \phi - \cot{\chi} \partial_\chi \phi \right) = 0
\end{equation}
The solution can be separated, $\phi(\rho,\chi) = f(\rho) g(\chi)$ and
the resulting equations solved in terms of a separation constant $k$,
which we may chose to be positive. The radial equation is easily
solved giving,
\begin{equation}
f(\rho) = \rho^{\pm k - 1}
\label{eq:fsoln}
\end{equation}
and since we wish the back-reaction to be small, we may only take
solutions that are regular as $\rho \rightarrow 0$.

Now we consider the angular equation. The horizon, at $\chi = 0, \pi$,
must be an equipotential surface.  Due to the separable form, we see
this may either be achieved by $f$ being constant (ie. k = 1), or by
$g = 0$ at $\chi = 0, \pi$. The first case yields only the trivial
solution $\phi = const$.  The second is more subtle, and we find the
solution,
\begin{equation}
g(\chi) = \sin^2{\chi} \, F\left(\frac{3-k}{4}, \frac{3+k}{4}; 2; \sin^2{\chi} \right)
\end{equation}
where $F(a,b;c;x)$ is the hyper-geometric function. Additionally
we add the geometric boundary condition at $\chi= \pi/2$, where
the differential equation is singular, that $g(\chi)$ be
reflection symmetric. The hyper-geometric function is a finite
degree polynomial in $\sin^2{\chi}$ if $k = 3 + 4 n$ with integer
$n \ge 0$, and hence has the required reflection symmetry.  Indeed
we may Taylor expand $g$ about $\chi = \pi/2$ giving,
\begin{equation}
g(\chi) = \frac{\sqrt{\pi}}{\Gamma{\left(\frac{5-k}{4}\right)} \Gamma{\left(\frac{5+k}{4}\right)}} - \frac{2 \sqrt{\pi}}{\Gamma{\left(\frac{3-k}{4}\right)} \Gamma{\left(\frac{3+k}{4}\right)}} \cos{\chi} + O(\cos{\chi}^2)
\end{equation}
which confirms that only for these values of $k$ is the reflection
boundary condition obeyed.

If we consider the exact cone geometry these modes with $k = 3 + 4 n$
clearly blow up at large $\rho$, far from the apex.  However, for the
non-uniform string, it is unclear what the geometry will be for large
$\rho$ and therefore we cannot specify precisely the remaining
boundary condition for the radial equation. Instead we assume that the
leading order term is not absent in the solution, so the dominant
contribution to the potential near the apex is from the lowest $k$, $k
= 3$, giving,
\begin{equation}
\phi = q \left( \rho \sin{\chi} \right)^2 + O(\rho^6)
\label{eq:phisoln}
\end{equation}
for constant $q$, the higher corrections coming from $n \ge 1$.
Therefore we see the electric field vanishes on the cone horizon
confirming our expectation of a reduced field near the apex.

To conclude, we still expect the cone to describe the limiting waist
geometry for a weakly charged string.  Obviously it would be
interesting to understand how this picture changes as the charge is
increased to extremality where one expects the GL instability to
disappear \cite{GL3, GL4, HM2}.

%==============================================================================
%
\section*{Acknowledgements}
%
%==============================================================================

We would like to thank Adam Ritz for useful discussions.  The work of
BK is supported in part by the Israeli Science Foundation. TW is
supported by Pembroke College, Cambridge, and would like to thank YITP
for hospitality during completion of this work.  Computations were
performed on COSMOS at the National Cosmology Supercomputing Centre in
Cambridge.

%==============================================================================
%
\newpage \bibliography{cone}
%
%==============================================================================

%==============================================================================
%
\end{document}